# High growth rate 4H-SiC epitaxial growth using dichlorosilane in a hot-wall CVD reactor

Iftekhar Chowdhury[1], MVS Chandrasekhar[1], Paul B Klein[2], Joshua D. Caldwell[2] & Tangali Sudarshan[1].

[1]University of South Carolina, Electrical Engineering, Columbia, SC 29208
[2]Naval Research Laboratory, Washington D.C 20375, USA

**Abstract**

Thick, high quality 4H-SiC epilayers have been grown in a vertical hot-wall chemical vapor deposition system at a high growth rate on (0001) $8^0$ off-axis substrates. We discuss the use of dichlorosilane as the Si-precursor for 4H-SiC epitaxial growth as it provides the most direct decomposition route into $SiCl_2$, which is the predominant growth species in chlorinated chemistries. A specular surface morphology was attained by limiting the hydrogen etch rate until the system was equilibrated at the desired growth temperature. The RMS roughness of the grown films ranged from 0.5-2.0 nm with very few morphological defects (carrots, triangular defects, etc.) being introduced, while enabling growth rates of 30-100 μm/hr, 5-15 times higher than most conventional growths. Site-competition epitaxy was observed over a wide range of C/Si ratios, with doping concentrations $< 1 \times 10^{14}$ $cm^{-3}$ being recorded. X-ray rocking curves indicated that the epilayers were of high crystallinity, with linewidths as narrow as 7.8 arcsec being observed, while microwave photoconductive decay (μPCD) measurements indicated that these films had high injection (ambipolar) carrier lifetimes in the range of 2 μs.

# Introduction

Silicon carbide is a desirable material for high power and high frequency devices due to its wide band gap, high break-down field and high thermal conductivity compared to silicon. Furthermore, the higher junction operating temperatures possible with SiC (~400°C), in comparison to silicon (~100°C), reduces the cooling requirements of SiC-based power systems, allowing compact, high performance modules to be realized.

In recent years, there has been strong interest in high-power devices with blocking voltages in excess 10 kV. Several papers have been published on power DMOSFETs [1], implanted VJFETs [2], PiN diodes [3], and Schottky diodes [4]. In all these devices, an epitaxial layer of 80-100 μm is required to achieve large blocking voltages. To obtain such thicknesses with standard silane-based chemical vapor deposition (CVD) processes, which have typical growth rates of 6-7 um/hr [5], process times would exceed ten hours, leading to a significant increase in the manufacturing cost. Therefore, the development of SiC epitaxy with growth rates exceeding 50 μm/hr is highly desired.

Commercial SiC CVD processes typically use silane and light hydrocarbons, such as propane or ethylene, diluted in hydrogen as a carrier gas. While growth rates higher than the usual 6-7 um/hr [5] may be achieved by increasing precursor flow, this typically leads to the homogeneous nucleation of liquid silicon droplets, which react with the carbon precursors to generate an encapsulating SiC layer over the Si core [6]. Once encapsulated, the evaporation of these Si droplets is no longer possible. Such SiC-coated droplets eventually decrease the efficiency of precursor use and degrade crystal quality.

Halide-based precursors have been demonstrated by several research groups to overcome problems with limited growth rate and Si-droplet formation [7, 8]. However, in this paper we report for the first time an epitaxial chemistry based on dichlorosilane as a Si-precursor. Under the growth conditions discussed in these reports, experiments showed that the occurrence of silicon droplets at the growth front were reduced due to the presence of halogens that have a stronger affinity for silicon than silicon itself. The average bond enthalpies at $25^0$C for the Si-Si, Si-F, Si-Cl, Si-Br, and Si-I bonds are 226, 597, 400, 330, and 234 kJ/mol [9], respectively. Due to the large size of bromine and iodine atoms, these species have relatively weak bonds with Si, whereas the much smaller fluorine atoms interact with Si too strongly, making their dissociation for growth inefficient. Thus, chlorine remains as the best compromise for halide-based epitaxial growth chemistries. Chlorine-based growth chemistries are initiated in the CVD reactor during standard SiC growths as either HCl [5, 7, 10, 11], $SiH_xCl_y$ [12-14], $CH_xCl_y$ [8], or $SiC_xH_yCl_z$ [15] molecules. Several theoretical and experimental studies investigating chlorine-based epitaxy have been made and growth rates exceeding 110 µm/hr [5] has been shown. The primary precursors discussed in these studies are tetra-, tri-, and dichlorosilane. Thermal dissociation of these precursors leads to the formation of $SiCl_2$ [16], which is the primary growth species. The corresponding gas phase reactions and associated activation energies for these precursors are presented in Table 1.

From Table 1, it would appear that in comparison to dichlorosilane, the decomposition of trichlorosilane into $SiCl_2$ proceeds with reduced activation energy of 4 kcal/mol (77.4-73.7 kcal/mol). However, this does not imply a more favorable decomposition reaction into $SiCl_2$ as the primary decomposition reaction for trichlorosilane provides $SiCl_3$ as a product, as this process requires only 2.5 kcal/mol. Therefore, trichlorosilane decomposition predominantly leads to a surplus of $SiCl_3$; this process is an inefficient route to the formation of $SiCl_2$ which is the primary growth species. Based on this, the use of dichlorosilane can be expected to provide optimal results due to its single gas phase decomposition reaction (Table 1) and comparable activation energy. Further, dichlorosilane is commercially available in the gas-phase at room temperature, which removes the need for a bubbler, thereby simplifying the reactant injection process.

**Experimental**

The epitaxial films discussed here were all grown in a home-built vertical hot-wall reactor that allows for SiC growth on substrates up to 2 inches in diameter in a chimney configuration [19]. The chamber was designed to reach temperatures up to 2200 $^0$C using graphite insulation. The substrates used were all research grade Si-face, n-type ($N_d \approx$ 1x10$^{18}$ cm$^{-3}$), 4H-SiC (0001) oriented 8$^0$ off-axis towards the [11-20] direction. In this study, we have used 8 mm x 8 mm square samples.

Control samples were grown with standard precursor chemistries (silane and propane) in a hydrogen atmosphere [20]. These growths were carried out at 1550$^0$C using 3 sccm silane and 0.9 sccm propane at 300 torr. Growth rates up to 15 μm/hr were observed, although evidence of significant formation of silicon droplets and comet-like inclusions (Figure 1) was observed at the highest growth rates. Further experiments over a range of growth temperatures did not lead to significant improvements in either the growth rate or the surface morphology. Figure 1 shows the surface morphology of such an epilayer grown at ~15 μm/hr for 1 hour.

For the test samples discussed here, dichlorosilane and propane were used as precursor gases with hydrogen used as the carrier gas. The hydrogen flow rate was varied from 6-12 slm. The flow rates of dichlorosilane and propane were 3.5-5.3 sccm and 1-2 sccm, respectively. The C/Si ratio was varied in the range of 0.89-1.7, while the growth temperature ranged from 1650-1750$^0$ C at 300 torr. Typical growth times were 30-60 minutes, with growth rates of ~50-75 μm/hr typically being observed, although a maximum growth rate of 108 μm/hr was obtained. The epilayers were unintentionally doped by varying the C/Si according to the site competition epitaxy rule [21], with doping concentrations of the as-grown epilayers ranging from $N_d=2\times10^{14}$ cm$^{-3}$ to $5\times10^{15}$ cm$^{-3}$. In total, this study was comprised of over 50 growths.

At the start of the epitaxial growth process the substrate was loaded into the reactor and the system was pumped down to a pressure of P≈$1\times10^{-7}$ torr. Several argon purge cycles were carried out during the process to assist in the removal of nitrogen from the graphite crucible. A 1000$^0$C bake out was carried out under vacuum to reduce the concentration of background impurities [22].

Figure 2 shows the temperature profile for a typical 4H-SiC growth using dichlorosilane. The growth was initiated with a hydrogen etch sequence once the system equilibrated at the prescribed growth temperature (1650-1750$^0$ C). The precursors were then introduced into the hydrogen atmosphere and the growth was initiated at a fixed pressure of 300 torr. The surface morphology of the epilayer was imaged post-growth using Nomarski diffractional interference contrast imaging. The epitaxial layers were analyzed by anodic oxidation or Fourier transform infrared reflectance (FTIR) to determine the film thickness, by mercury-probe C-V measurements to verify the doping density, AFM for surface roughness analysis, and molten KOH etching to quantify the density of extended defects. Carrier lifetimes were subsequently measured by both microwave photoconductive decay (µPCD) and time-resolved photoluminescence (TRPL), while micro-Raman spectroscopy and X-ray diffraction rocking curves were used to quantify the crystalline quality.

**Results and Discussion**

**Effect of hydrogen flow**

To determine the influence of hydrogen carrier gas flow rate upon the epitaxial growth, the growth rate and RMS roughness of the epilayers were recorded as a function of the carrier gas flow rate. These experiments were conducted at $1750^0$ C at 300 torr using 3.85 sccm dichlorosilane and 1.1 sccm propane, while the hydrogen flow rate was varied from 6 slm -12 slm. The dependence of RMS roughness and growth rate upon the hydrogen flow is shown in Figure 3. We will discuss this trend in the context of hydrogen etching, precursor partial pressure, and the boundary layer [23, 24] assumption.

As shown in Figure 3(a), a large number of droplets [5] were formed and deposited on the 58 µm thick epilayer that was grown at 6 slm of hydrogen flow, and AFM measurements indicated that this film had the highest RMS roughness of the samples grown within this study at 5.16 nm. As the hydrogen flow was increased to 8 slm [Figure 3(b)], the RMS roughness improved to ~2.0 nm, and the growth rate dropped to 48 µm/hr. This trend continued with further increases in the hydrogen flow; the RMS roughness of the film grown at 12 slm [Figure 3(c)] improved further dropping to 0.55 nm. These results can be understood as follows. The increased hydrogen flow rate increases the etching rate, thereby decreasing the surface roughness, which is consistent with the observations of Kumagawa et al. [25].

This increased etch rate at higher hydrogen flows can be explained by the boundary layer assumption. Boundary layer is a stagnant layer of gas present at solid-gas interfaces in the presence of laminar gas flow. As the hydrogen flow increases from 6 slm to 12 slm, the boundary layer gets thinner [23]. Furthermore, increased hydrogen flow leads to decreases in the partial pressures of carbonaceous species, which are the byproduct of hydrogen etching of SiC. Thus, these species diffuse out through the thin boundary layer more efficiently, thereby increasing the etch rate, which in turn results in an improved surface roughness.

In Figure 3 we also observe that increasing the hydrogen flow decreases the growth rate. If we are to consider only precursor dilution by increased hydrogen flow, the growth rate should have decreased linearly to 29 µm/hr for 12 slm of hydrogen flow. However, the growth rate only decreased to 42 µm/hr. The excess growth rate at 12 slm may be due to one, or both of the following reasons. The thinning of the boundary layer, which enhances precursor transport to the SiC surface, counteracting the influence of reduced precursor partial pressure. The other possible reason is that low $H_2$ flow rates can cause precursor depletion [26]. This depletion becomes less severe when $H_2$ flow rate is increased which explains the excess growth rate at 12 slm $H_2$ flow.

**Effect of precursor flow**

An additional set of experiments were completed to verify that the growth regime was transport limited. The propane flow was fixed at 1.8 sccm, while the dichlorosilane flow was increased from 3.5 sccm to 4.9 sccm, consequently increasing the dichlorosilane partial pressure.

Figure 4(a) shows that the growth rate increases with increasing dichlorosilane flow, which is consistent with SiC epitaxy being limited by mass transport of the Si-species. However, there is an upper boundary of dichlorosilane flow (4.9 sccm) [shown by dashed line in Figure 4(a)], above which Si-clusters begin to nucleate in the vapor phase, which in turn deposit as droplets on the surface. This limited the viable growth rate to ~50 µm/hr. Increasing propane flow rate at this point does not help in reducing droplet formation. Also, growth rate does not increase.

To reduce the Si-cluster formation and thus depletion of precursor that is promoted at higher dichlorosilane flow rates, the hydrogen flow rate was increased to reduce the dichlorosilane partial pressure. This approach leads to a C-limited regime [Figure 4(b)] enabling the realization of specular surfaces while increasing the growth rate to as high as 100 µm/hr. This effect was also observed in a previous study [27] using a hot-wall chimney reactor and silane and propane as precursors, where the growth was found to be Si-limited for C/Si ~ 1.0, whereas in our system such ratios led to a C-limited regime.

This demonstrates that the use of dichlorosilane helped to increase the Si species concentration at the growth front without introducing homogeneous Si droplet nucleation. By optimizing our growth conditions, the Si-transport barrier was broken; leading to a C-limited growth regime, thereby enabling growth rates exceeding 100 µm/hr while maintaining a specular surface.

**Effect of C/Si ratio**:

The C/Si ratio of the inlet gas mixture is an important process parameter in the CVD growth of SiC. It is known to influence the incorporation of impurities, as well as surface morphology [28]. The "site-competition theory" presented by Larkin [29] showed that nitrogen incorporates into the carbon-site, while aluminum displaces silicon in the SiC lattice. The site competition epitaxial technique is based on appropriately controlling the C/Si ratio in the reactor in an effort to effectively control the efficiency of impurity incorporation at the substitutional SiC lattice sites. As nitrogen and carbon compete for the C-site and aluminum, boron and silicon compete for the Si-site, C-rich conditions lead to p-type and Si-rich conditions to n-type conductivity.

To study how the C/Si ratio impacts both the growth morphology and doping density, experiments with C/Si ratios varying from 0.81 to 1.7 were performed. In these experiments, the DCS flow rate was maintained at a fixed value, while propane flow was varied to modify the C/Si ratio. The hydrogen flow was held constant at 12 slm and no intentional dopant was introduced.

As shown in Figure 5, epilayers grown with C/Si > 1.4 were found to be p-type and the doping density increased with larger C/Si ratios, while epilayers grown with C/Si <1.4 were found to be n-type. This doping dependence behavior was also previously reported for other halide chemistries [15] and silane-propane epitaxy. Doping concentrations $< 1\times10^{14} cm^{-3}$ were achieved in this investigation.

Figure 6 illustrates that the RMS roughness decreased from 2.04 nm to 0.47 nm as the C/Si ratio was increased from 0.81 to 0.89, with this reduction in surface roughness being coupled with a significant reduction in morphological defects, as observed by Nomarski microscopy (Figure 6). However, further increases in the C/Si ratio to 1.7 led to the nucleation of many triangular defects and a corresponding increase in the RMS roughness.

**Raman Spectroscopy**

Raman scattered light was collected using a Jobin Yvon spectrometer and was focused onto a charged- coupled-detector (CCD) array for detection. The excitation was carried out at room temperature using a 25 mW, 632 nm line of a He-Ne laser. The laser beam was focused onto the sample surface using an 80x, 0.75 N.A. confocal microscope objective, which also served to collect the backscattered light.

In the backscattering geometry, the Raman spectrum of (0001) 4H-SiC presents three main phonon bands with $A_1$, $E_1$ and $E_2$ symmetries. The planar $E_2$ transverse optic (TO) mode is observed at 776 cm$^{-1}$, the $E_1$ TO mode at 796 cm$^{-1}$ and the $A_1$ longitudinal optic (LO) mode at 964 cm$^{-1}$ [30]. Figure 7 compares typical Raman spectra averaged over a 40 μm x 40 μm area from a standard n$^+$ ($N_d \approx 1 \times 10^{19}$ cm$^{-3}$) 4H-SiC substrate and a 30 μm thick n$^-$ ($1 \times 10^{15}$ cm$^{-3}$) epilayer grown on a similar substrate using the dichlorosilane based growth described here, at room temperature.. As shown in Table 2, the linewidth of the $E_2$ TO peak for the epilayer was 3.75 cm$^{-1}$, while the substrate was measured to be 4.5 cm$^{-1}$ thereby illustrating an improvement in the crystalline quality of the epilayer.

Nakashima et al [30] showed that 3C-SiC has a single, dominant $E_1$(TO) mode that is also located at 796 cm$^{-1}$. Therefore, the ratio $E_2$(TO)/$E_1$(TO) for a 4H-SiC sample provides some insight into the presence and location of 3C polytype inclusions, which are a common defect in 4H-SiC epilayers. Since the substrate and the epilayer have similar linewidths for each peak, a reduction in the $E_2(TO)/E_1(TO)$ ratio implies a higher probability of 3C inclusions in the grown epilayer. The ratio for these samples was typically on the order of 60 for the epilayers, whereas a value of ~32 was found for the substrates (Table 2), which clearly shows that the epitaxial growth did not introduce or expand any regions of 3C-SiC inclusions that may have been present in the substrate.

**X-ray diffraction rocking curve:**

X-ray diffraction was performed to measure the crystalline uniformity of the 4H-SiC films. A rocking curve of the (0008) X-ray diffraction peak for a typical 35 μm thick epilayer is shown in Figure 8, FWHM is ~ 9 arcsec. Overall the FWHM for all of the samples was found to fall between 7.8 – 14.1 arcsec, with the theoretical lower limit for SiC (0001) crystals being approximately ~6-8 arcsec [20], thereby indicating that our epilayers were high quality monocrystalline films. Other epilayers studied in this work exhibited similar narrow rocking curves. In comparison with other chlorosilane [15, 31] work, the low FWHM value of our epilayers demonstrates excellent crystal quality as indicated in Figure 8.

In an effort to quantify the density of extended defects within the epilayers, molten KOH etching was performed at $650^0C$. From these efforts an average epilayer etch-pit density (extended defect density) was found to be approximately $1.2 \times 10^3$ cm$^{-2}$, whereas the average density found in the commercial substrates was around $5 \times 10^4$ cm$^{-2}$, once again confirming the high quality of these films produced at high growth rates [32].

**Carrier lifetime:**

Carrier lifetimes in our epilayers were measured using two optical techniques, Microwave($\mu$)-PCD and TRPL, both of which are nondestructive and require no device structure. $\mu$-PCD measurements were carried out at room temperature with a Semilab WT-85 lifetime scanner that was modified to include the 355 nm output from a 3XNd:YAG laser with a pulse width of 50 ns (<15 ns fall time) and an average power of ~200 mW. The system was upgraded to improve the temporal resolution to 10 ns per point. TRPL measurements were also carried out at room temperature, while the excitation was provided by a frequency-doubled, mode-locked and cavity-dumped Ti:sapphire laser (355 nm, 150 fs pulse width, 100-500kHz, 5 nJ/pulse). The average injection level over the layer was $\approx 2 \times 10^{14} cm^{-3}$. Figure 9 shows TRPL transients collected from samples 1 and 2 under low injection level.

In Figure 9, the fast components (20-30 ns) contain contributions from surface recombination, background radiation and substrate emission, whereas the slow component is representative of the minority carrier lifetime $\tau_{MCL}$. It should also be noted that the lifetime of sample 1 (759 ns at 60 $\mu$m thickness) and 2 (742 ns at 30 $\mu$m thickness) are comparable, even with a factor of two difference in thickness.

$\mu$-PCD lifetime measurements were performed on the same sample for direct comparison. Taking into account the high injection conditions (~$1 \times 10^{16}$ cm$^{-3}$) and variations in the carrier dynamics monitored by the two methods, the carrier lifetimes measured via the $\mu$-PCD technique in sample 1 (2.06 $\mu$s) and 2 (2.35 $\mu$s) are consistent [33] with the corresponding TRPL lifetimes. Under such high injection conditions, the $\mu$-PCD lifetime may be referred to as the ambipolar lifetime.

A map of the carrier lifetime for sample 1 is presented in Figure 10 (a). Other than the edge exclusion (~1 mm) the blue area shows uniform lifetime over the whole sample. Figure 10(b) shows the power dependence of lifetime for the same sample measured by µ-PCD. Here, increasing laser power (increasing injection) was found to cause a reduction in the carrier lifetime. Such a power dependence was reported by Kimoto [34] et al for samples with a low density of deep-level electron traps ($<\sim 10^{13\text{-}14}$ cm$^{-3}$), once again implying that the films used in this study were of high quality.

Table-2 Summarizes the quantitative information recorded on various epitaxial layers and substrates using the characterization techniques discussed in the text.

**Conclusions**

A high growth rate (up to 100 µm/hr) was attained for 4H-SiC epitaxial growth and was demonstrated in a chimney-type hot-wall CVD reactor using dichlorosilane, a precursor expected to have superior cracking kinetics in halide-assisted SiC epitaxy. The growth process can be divided into two regimes, where the system is either Si- or C-transport limited. By increasing the C/Si ratio at high growth rates in the presence of dichlorosilane, a C-limited regime was achieved. This enabled the elimination of Si-cluster formation, leading to specular surface morphologies over a wide range of C/Si ratios, with epilayer thicknesses up to ~100 µm being grown. The as-grown epilayers were characterized via XRD and Raman scattering to be of high crystalline quality with no observable in-grown 3C-SiC inclusions or other morphological defects and exhibited ambipolar lifetimes in excess of 2 µs.

**Acknowledgements**

This work was supported by the Office of Naval Research, Grant no. N000140910619. The authors thank contract monitor Dr. H. Scott Coombe of ONR for his support of this research. The authors also thank Dr. Chris Williams for the use of his Raman Spectroscopy setup and Dr. Peter Muzykov for his assistance in the XRD measurements.

Table 1- Thermal dissociation of di-, tri-, and tetrachlorosilane and the corresponding gas phase reactions and associated activation energies

| Reaction | E (kcal/mol) | Ref. |
|---|---|---|
| **SiH$_2$Cl$_2$** | | |
| $SiH_2Cl_2 \leftrightarrow SiCl_2 + H_2$ | 77.4 | [17] |
| **SiHCl$_3$** | | |
| $SiHCl_3 + H \leftrightarrow SiCl_3 + H_2$ | 2.5 | [18] |
| $SiHCl_3 \leftrightarrow SiCl_2 + HCl$ | 73.7 | [17] |
| **SiCl$_4$** | | |
| $SiCl_4 \leftrightarrow Cl + SiCl_3$ | 111.16 | [16] |
| $SiCl_3 + H_2 \leftrightarrow SiHCl_3 + H$ | 2.5 | [18] |
| $SiHCl_3 \leftrightarrow SiCl_2 + HCl$ | 73.7 | [17] |

Table 2- Summary of the quantitative information recorded on various epitaxial layers and substrates using the characterization techniques discussed in the text.

|  | Substrate | Epilayer |
|---|---|---|
| X-ray rocking curve (FWHM) (arcsec.) | 20 ~ 40 | 7.8 - 14.1 |
| Raman, $E_2$ (TO) FWHM (cm$^{-1}$) | 4.5 | 3.75 |
| $\frac{E_2(TO)}{E_1(TO)}$ ; (3C inclusion) | 32 | 60 |
| Etch pit density (cm$^{-2}$) | ~5 x10$^4$ | ~1.2 x10$^3$ |
| µ-PCD Lifetime (µs) | ~0.1 | 2.0 |
| TRPL(µs) | 0.03 | 0.8 |

**Figure Captions**

Figure 1. Nomarski optical microscopy of a 15 μm thick epilayer grown on a Si-face, n-type ($N_d \approx 1\times10^{18}$ cm$^{-3}$), 4H-SiC (0001), $8^0$ off-axis substrate using silane and propane.

Figure 2. Temperature profile for the epilayer growth process.

Figure 3. AFM and Nomarski images taken of 58, 48 and 42 μm thick undoped epilayers grown with different hydrogen flow rates (C/Si = 0.89) at $1750^0$ for 1 hour. The surface roughness is indicated in the images.

Figure 4. (a) Dependence of growth rate on DCS flow rate (b) Dependence of growth rate on Propane flow rate.

Figure 5. C/Si ratio dependence of the doping concentration for the unintentionally doped layers.

Figure 6. AFM and Nomarski images taken from undoped epilayers each 52μm thick grown with different C/Si ratios at $1750^0$C. Surface roughness is indicated in the images.

Figure 7. Raman spectra of n-type 4H-SiC epilayer (1E+15 cm$^{-3}$) at room temperature in backscattering geometry.

Figure 8. High resolution X-ray diffraction rocking curve of a 35 μm thick epilayer grown on a 4H-SiC, 8 degree off towards (11-20) substrate.

Figure 9. TRPL transient of Sample 1 and Sample 2 with thickness of 60 μm and 30 μm respectively, at room temperature. Doping concentration around $2\times10^{14}$ cm$^{-3}$.

Figure 10. (a) Spatial map of the carrier lifetime under high injection conditions (~$1\times10^{16}$ cm$^{-3}$) using the μ-PCD technique (b) Power dependence of the carrier lifetime near the center of sample 1.

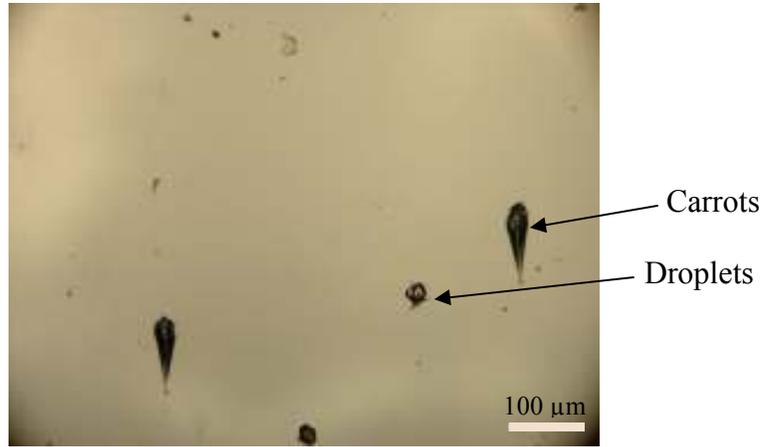

Figure 1.

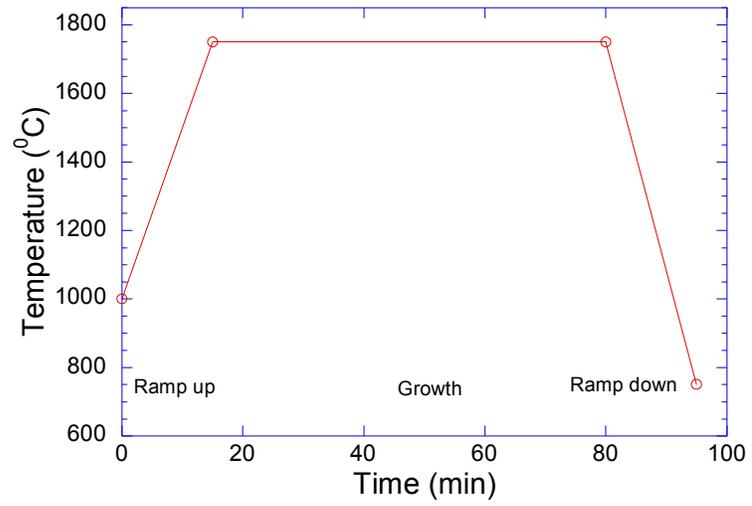

Figure 2.

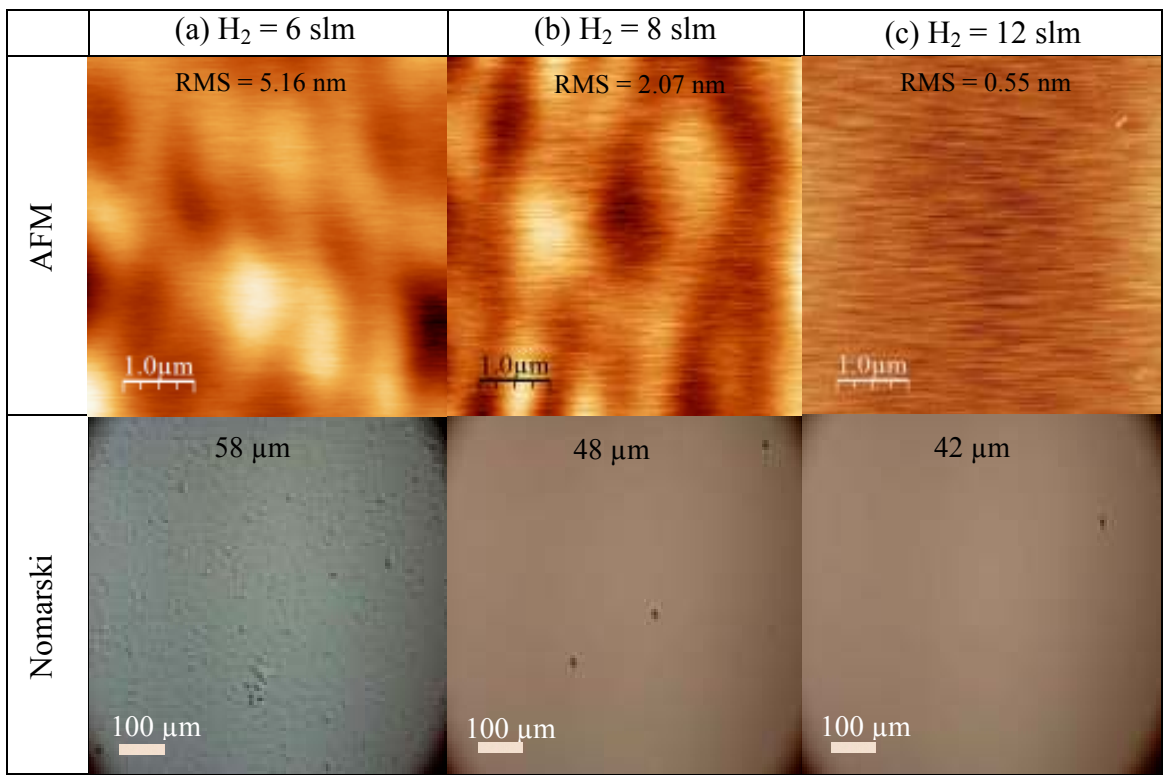

Figure 3.

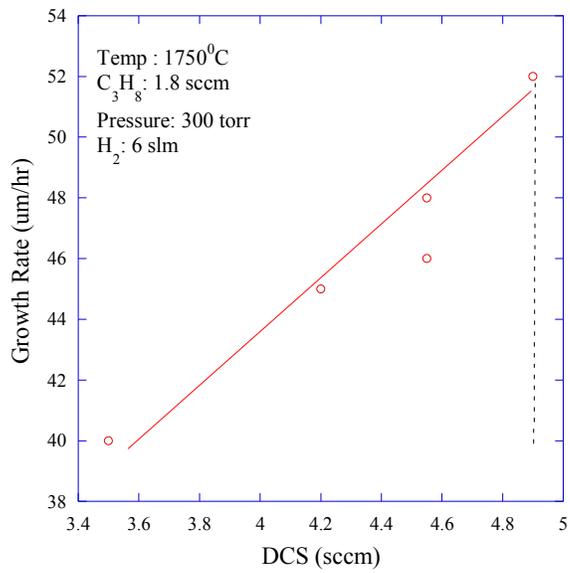

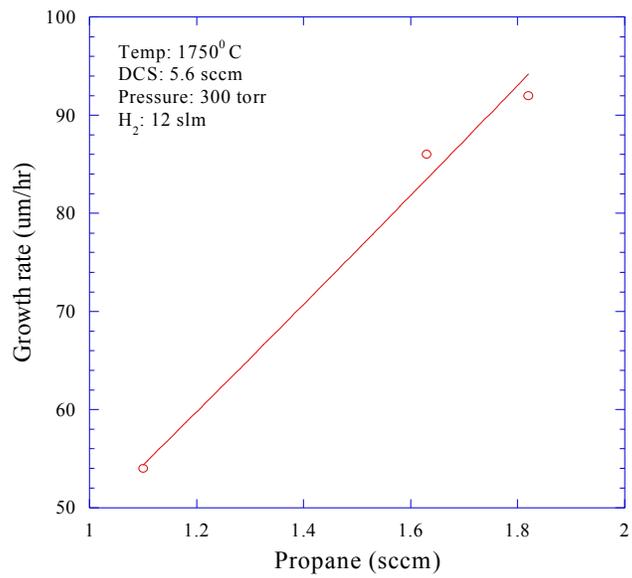

Figure 4.

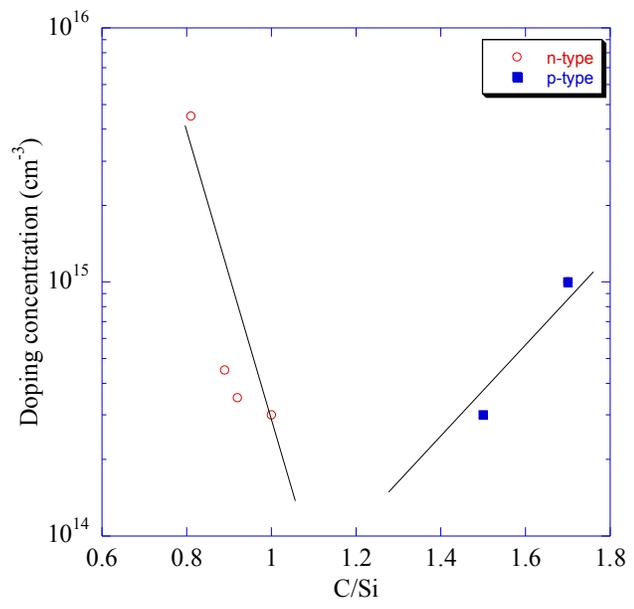

Figure 5.

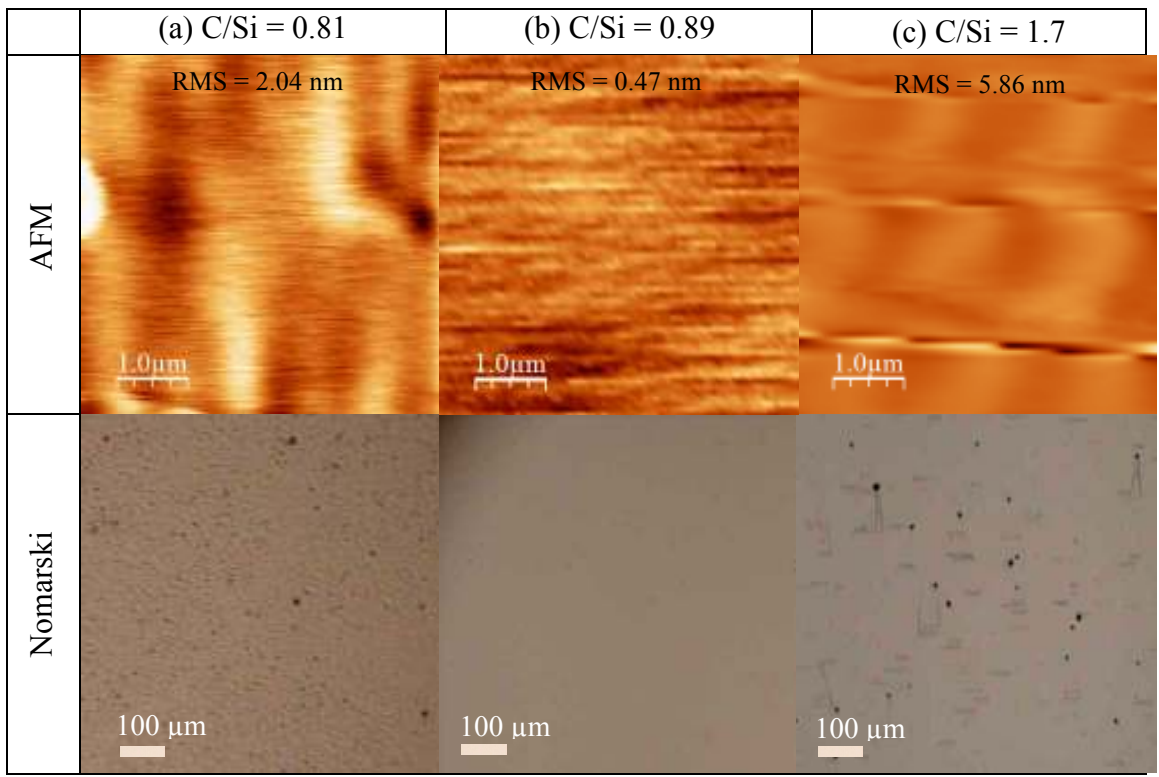

Figure 6.

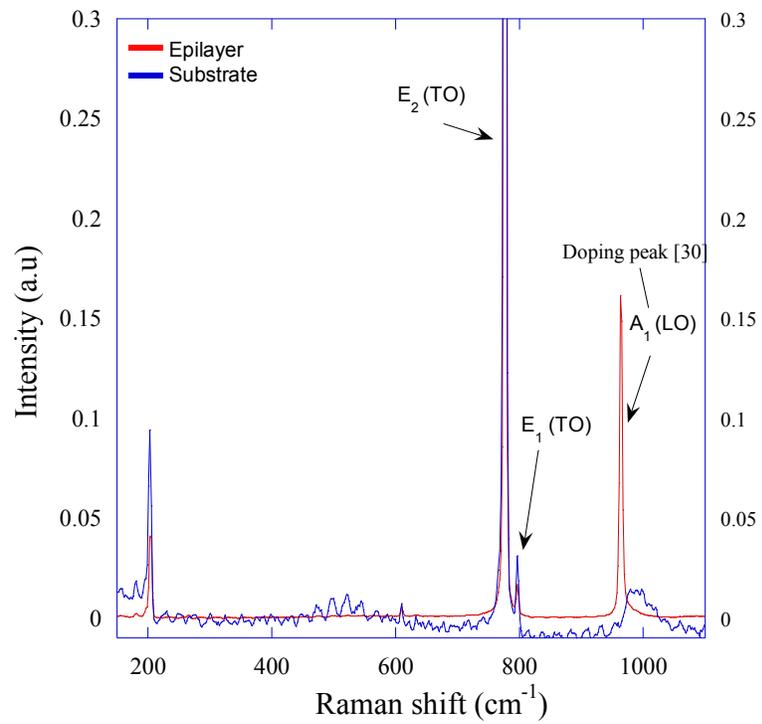

Figure 7.

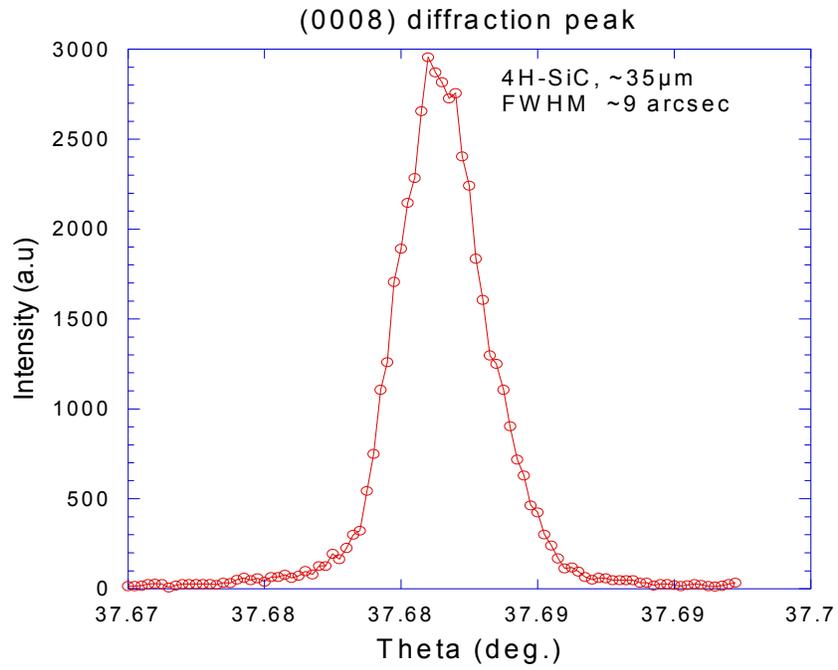

| | H Pedersen[15] | S Nakazawa[31] | USC |
|---|---|---|---|
| FWHM (arcsec) | 10 | 9 | 9 |

Figure 8.

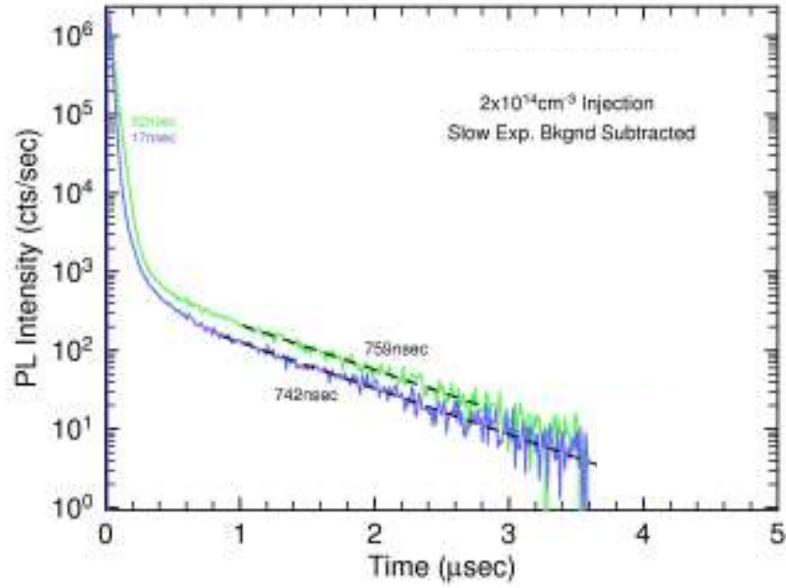

Figure 9.

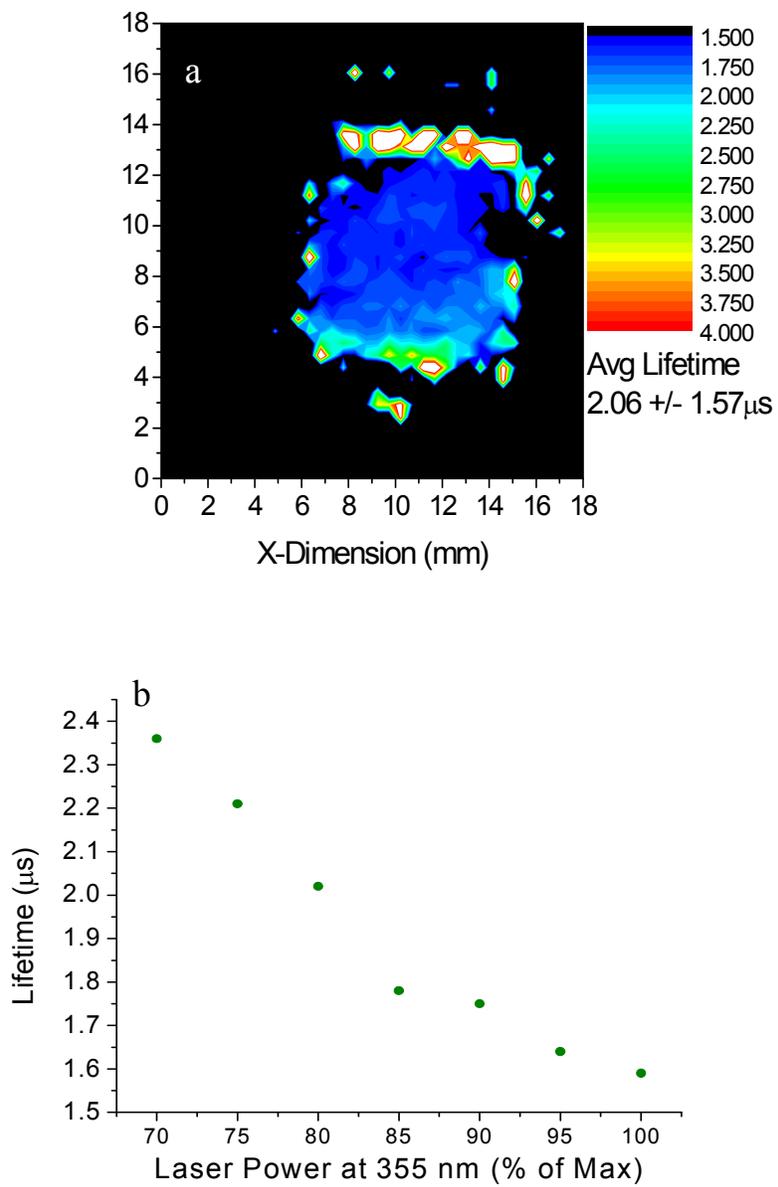

Figure 10.